\documentclass[fleqn,usenatbib]{mnras}

\usepackage{newtxtext,newtxmath}

\usepackage[T1]{fontenc}
\usepackage{ae,aecompl}

\usepackage{graphicx}	
\usepackage{amsmath}	
\usepackage{amssymb}	
\usepackage{physics}
\usepackage{tabularx}

\newcommand\T{\rule{0pt}{2.6ex}}       
\newcommand\B{\rule[-1.2ex]{0pt}{0pt}} 

\title[Stellar magnetic helicity density]{Measuring stellar magnetic helicity density}

\author[K. Lund et al.]{K. Lund$^{1}$\thanks{E-mail: kblg@st-andrews.ac.uk}, 
M. Jardine$^{1}$, L. T. Lehmann$^{1}$, D. H. Mackay$^{2}$, V. See$^{3}$, A. A. Vidotto$^{4}$,
\newauthor
J.-F. Donati$^{5}$, R. Fares$^{6}$, C. P. Folsom$^{5}$, S. V. Jeffers$^{7}$, S. C. Marsden$^{8}$, J. Morin$^{9}$
\newauthor
and P. Petit$^{5}$ 
\\
$^{1}$SUPA, School of Physics and Astronomy, University of St Andrews, North Haugh, St Andrews, KY16 9SS, UK\\
$^{2}$School of Mathematics and Statistics, University of St Andrews, North Haugh, St Andrews, KY16 9SS, UK\\
$^{3}$University of Exeter, Department of Physics \& Astronomy, Stocker Road, Devon, Exeter, EX4 4QL, UK\\
$^{4}$School of Physics, Trinity College Dublin, the University of Dublin, College Green, Dublin-2, Ireland\\
$^{5}$IRAP, Universit\'{e} de Toulouse, CNRS, UPS, CNES, 14 Avenue Edouard Belin, 31400, Toulouse, France\\
$^{6}$Physics Department, United Arab Emirates University, P.O. Box 15551, Al-Ain, United Arab Emirates\\
$^{7}$Institut f\"ur Astrophysik, Universit\"at G\"ottingen, Friedrich-Hund-Platz 1, D-37077 G\"ottingen, Germany\\
$^{8}$University of Southern Queensland, Centre for Astrophysics, Toowoomba, QLD, 4350, Australia\\
$^{9}$LUPM, Universit\'e de Montpellier, CNRS, Place Eug\`ene Bataillon, F-34095 Montpellier, France
}

\date{Accepted XXX. Received YYY; in original form ZZZ}

\pubyear{2019}

\begin{document}
\label{firstpage}
\pagerange{\pageref{firstpage}--\pageref{lastpage}}
\maketitle

\begin{abstract}
 
Helicity is a fundamental property of a magnetic field but to date it has only been possible to observe its evolution in one star - the Sun. In this paper we provide a simple technique for mapping the large-scale helicity density across the surface of any star using only observable quantities: the poloidal and toroidal magnetic field components (which can be determined from Zeeman-Doppler imaging) and the stellar radius.  We use a sample of 51 stars across a mass range of 0.1-1.34 M$_\odot$ to show how the helicity density relates to stellar mass, Rossby number, magnetic energy and age. We find that the large-scale helicity density increases with decreasing Rossby number $R_o$, peaking at $R_o \simeq 0.1$, with a saturation or decrease below that. For both fully- and partially-convective stars we find that the mean absolute helicity density scales with the mean squared toroidal magnetic flux density according to the power law: $|\langle{h\,}\rangle|$ $\propto$ $\langle{\rm{B_{tor}}^2_{}\,\rangle}^{0.86\,\pm\,0.04}$. The scatter in this relation is consistent with the variation across a solar cycle, which we compute using simulations and observations across solar cycles 23 and 24 respectively. We find a significant decrease in helicity density with age.

\end{abstract}




\section{Introduction}

The helicity of a magnetic field is one of its most powerful measures. As an invariant of the ideal MHD equations \citep{Woltjer1958,Taylor1974}, it is a fundamental ingredient in our understanding of magnetic field generation and evolution. It relates the large and small scales within a magnetic field and constrains the evolution of that field towards a lowest-energy state. Helicity that has been captured by stars during their formation may be enhanced by the stellar dynamo and returned to the interstellar medium by the action of stellar winds and ejecta \citep{Berger2000,Zhang2005,Zhang2013,Blackman2015}. 

Measuring the helicity ($H$) of an astronomical magnetic field is, however, extremely challenging as it is inherently a three dimensional quantity that measures the linkage of fields. It can be defined in terms of the vector potential ($\boldsymbol{A}$) and the corresponding magnetic field ($\boldsymbol{B}=\curl{A}$) as \citep{Woltjer1958}:
\begin{equation}
H=\int \boldsymbol{A}\cdot \boldsymbol{B} \rm{dV},
\label{eq:Hdef}
\end{equation}
where V is volume. This reveals one of the challenges inherent in determining helicity - that it is defined only for a given gauge. The transformation $A \rightarrow A + \nabla \psi$ gives the same magnetic field $\boldsymbol{B}$ but a different helicity. This problem is resolved in a closed magnetic volume where the helicity is well defined, but if some flux penetrates the boundary, we can only define the helicity relative to a given field, normally chosen to be the potential field with the same boundary flux \citep{Berger1984}.

To date helicity has been measured for the Sun, but not other stars. There are many areas of current solar research where helicity plays a significant role and thus we limit our discussion to a few examples of its application. Magnetic helicity is central to the understanding of the evolution and generation of magnetic fields, i.e. solar dynamo theory \citep{Brandenburg2005,Chatterjee2011}, as well as to characterising the topology of coronal magnetic fields \citep{Berger1984}. Active regions on the Sun, where the magnetic fields are especially strong, frequently give rise to explosive events such as solar flares and coronal mass ejections (CMEs). \cite{Rust1994} proposed CMEs are a direct result of the conservation of helicity, necessary in order to remove excess helicity from the Sun. Subsequently helicity has been studied extensively as a diagnostic of solar eruptivity \citep{Zhang2006,Zhang2008,Zhang2012,Nindos2013}. Attempting to improve space weather predictions \cite{Pariat2017} linked the build-up of magnetic helicity to the likelihood of a solar eruption, and \cite{Hawkes2018} used helicity flux to predict overall solar activity. Studies such as these have motivated numerous authors to attempt measurements of magnetic helicity in the solar atmosphere, see reviews by \cite{Demoulin2007} and \cite{Demoulin2009}. 

Magnetic helicity in stellar research is not as well studied. Most recently, \citet{Warnecke2019} have modelled the impact of helicity on stellar X-ray emission, suggesting that the rise in X-ray emission with increasing rotation rate may be related to the underlying variation in the helicity.

In this paper we use observations of magnetic fields at stellar surfaces to extend the study of helicity from the Sun to a large sample of stars. Mapping magnetic fields across stellar surfaces has been made possible by the Zeeman-Doppler Imaging (ZDI) technique \citep{Semel1989} which has now been applied to a large range of stars (see, for example, \citet{Donati2009}). Many surveys, such as MagIcS\footnote{http://www.ast.obs-mip.fr/users/donati/magics/v1/}, Bcool\footnote{http://bcool.ast.obs- mip.fr/Bcool/Bcool\_\_\_cool\_magnetic\_stars.html}, MaTYSSE\footnote{https://matysse.irap.omp.eu/doku.php}, Toupies\footnote{http://ipag.osug.fr/Anr\_Toupies/} and MaPP\footnote{http://cfht.hawaii.edu/Science/MAPP/}, have explored magnetic field behaviour in a range of stars. One of the most intriguing early results was that for stars of similar mass to the Sun, field geometries are largely toroidal below Rossby number R$_o\approx$1 \citep{Donati2009}, prompting interest in exploring dynamo models that allow for strong surface toroidal fields \citep{Bonanno2016}. 

The ratio of toroidal to poloidal magnetic energies is well recovered by ZDI \citep{Lehmann2019} and appears to be different for stars that are fully convective and those that are only convective in the outer regions of their interiors \citep{Donati2008,Gregory2012,See2015}. Fully convective stars tend to be those that are either very young, or have low masses. Since they do not posses a tachocline at the interface between an inner radiative and outer convective zones, they cannot support a deep-seated interface dynamo. The nature of dynamo activity (and hence the types of field geometries produced) may therefore be different for these two different types of interior structure. A transition from strong, axisymmetric fields to weaker, more complex fields appears to occur in young stars when the radiative core forms (e.g. \citet{Donati2011,Folsom2016}). Studies of main sequence M dwarfs also show that for masses below $\sim$ 0.5 M$_\odot$ there is an apparent transition from weaker, complex fields to stronger, simpler fields \citep{Morin2008b,Morin2008c,Donati2009,Morin2010}. This may be related to a change in the nature of the dynamo across the fully convective boundary. 

\cite{Berger1985} and \cite{Berger2018} describe in detail how the poloidal-toroidal magnetic field decomposition allows for a simple expression of helicity as the net linking of toroidal and poloidal fields. This particular field decomposition has the added advantage of simplifying the treatment of the gauge. Since the corresponding potential field with the same boundary flux is purely poloidal and therefore has zero helicity, calculating the helicity in this decomposition is straightforward. 
 In the context of ZDI maps however, which only provide \textit{surface} magnetic fields, there is not enough information available to calculate magnetic helicities, as this is a property within a \textit{volume}. Consequently, in our work, we consider the surface magnetic helicity density instead. 
 
In a similar way \citet{Pipin2019} used poloidal and toroidal field components to calculate the evolution of the Sun's magnetic helicity density across solar cycle 24. They showed the relationship between the small-scale (active region) and large-scale (polar) fields through the solar cycle. According to their results, the large-scale and small-scale helicities started off with opposite signs at the beginning of the solar cycle, and then evolved to show the same sign in the declining phase. Furthermore, they measured the helicity of large-scale fields to be an order of magnitude smaller than the helicity of small-scale fields.

One of the limitations of the ZDI technique is that it is insensitive to small-scale magnetic flux elements whose polarities cancel out \citep{Reiners2009,Morin2010,Kochukhov2019,See2019}. We are therefore unable to explore the evolution of helicity across a large range of length scales in the way that is possible for the Sun. What we observe is rather the imprint of that helicity evolution on the largest scales. This may improve when moving to nIR data, e.g. using SPIRou, thanks to the larger Zeeman effect. By applying our technique to both observed and simulated solar magnetograms, however, we can place the Sun in the context of other stars and use this to help interpret stellar observations. We can determine the role of different length scales by expressing the poloidal and toroidal magnetic field components ($\boldsymbol{B}=\boldsymbol{B}_{\rm{pol}}+\boldsymbol{B}_{\rm{tor}}$) as a sum of spherical harmonics of different $l$-modes, where the smaller $l$-modes describe the larger-scale field and higher $l$-modes describe the smaller-scale field. Truncating the sum at some maximum $l$-value mimics the lack of sensitivity to small-scale fields.

In this paper we provide the reader with a simple equation for helicity density, given the poloidal and toroidal decomposition of any stellar magnetic field. We use this expression to study the large-scale helicity density of 51 stars. Our sample includes both fully- and partially-convective stars. Our aim is to discover how the helicity of these fields relates to fundamental stellar properties and to interpret these results in the context of the evolution of the large-scale helicity density of the Sun. 

The paper is outlined as follows: Section \ref{sec:Hderivation} describes the derivation of the magnetic helicity density in terms of poloidal and toroidal magnetic field components. In Section \ref{sec:Application} we calculate the helicity density of the Sun, as well as our sample of stars. Section \ref{sec:Discussion} presents our discussion of the results. A summary of our results along with our conclusions are given in Section \ref{sec:conclusion}.

\section{Magnetic helicity density}
\label{sec:Hderivation}

The magnetic helicity density is given by the integrand of Equation \ref{eq:Hdef}:
\begin{equation}
h=\boldsymbol{A}\cdot \boldsymbol{B}. 
\label{eq:hdef} 
\end{equation}
We expand the magnetic field and the vector potential field into their poloidal and toroidal components as follows: 
\begin{equation}
h=(\boldsymbol{A}_{\rm{pol}}+\boldsymbol{A}_{\rm{tor}})\cdot (\boldsymbol{B}_{\rm{tor}}+\boldsymbol{B}_{\rm{pol}}).
\label{eq:hpoltorfull} 
\end{equation}

An expression for the magnetic field is derived in Section \ref{sec:Bpoltor}, a corresponding vector potential is calculated in Section \ref{sec:Acalc}, and we combine them to obtain an expression for the helicity density in Section \ref{sec:Hcalc}. 

\subsection{Poloidal and toroidal magnetic field components}
\label{sec:Bpoltor}

We decompose the stellar magnetic field in terms of a poloidal and toroidal component following Appendix III of \cite{Chandrasekhar1961}:
\begin{equation}
\boldsymbol{B}_{\rm{pol}}=\curl{[\curl{[\Phi\hat{r}]}]},
\label{eq:Bporiginal}
\end{equation}
\begin{equation}
\boldsymbol{B}_{\rm{tor}}=\curl{[{\Psi\hat{r}}]}.
\label{eq:Btoriginal}
\end{equation}
In a spherical coordinate system\footnote{We use a spherical coordinate system where a positive radial field component points towards the observer, the meridional ($\theta$) component is positive pointing from North to South and the azimuthal ($\phi$) component is positive in the clockwise direction as viewed from the North pole.} the scalars $\Phi$ and $\Psi$ can be written in terms of spherical harmonics as
\begin{equation}
\Phi=S(r)c_{lm}P_{lm}e^{im\phi}
\label{eq:Phi}
\end{equation}
and
\begin{equation}
\Psi=T(r)c_{lm}P_{lm}e^{im\phi}.
\label{eq:Psi}
\end{equation}
$S(r)$ and $T(r)$ are functions describing the radial behaviour of the field, $P_{lm}$ is short for the Legendre Polynomial $P_{lm}(\cos\theta)$ of mode $l$ and order $m$, and $c_{lm}$ is the associated normalisation constant:
\begin{equation}
c_{lm}=\sqrt{\frac{2l+1}{4\pi}\frac{(l-m)!}{(l+m)!}}.
\label{eq:c}
\end{equation}

Expanding the poloidal and toroidal field components gives the following expressions: 
\begin{equation}
\begin{split}
B_{\rm{pol}}(r,\theta,\phi)&= \sum_{lm}\frac{l(l+1)}{r^2}S(r)c_{lm}P_{lm}e^{im\phi}\hat{r} \\
& +\sum_{lm}\frac{1}{r}\dv{S(r)}{r}c_{lm}\dv{P_{lm}}{\theta}e^{im\phi}\hat{\theta}\\
& +\sum_{lm}\frac{im}{r\sin\theta}\dv{S(r)}{r}c_{lm}P_{lm}e^{im\phi}\hat{\phi},
\label{eq:Bpexp} 
\end{split}
\end{equation}
\begin{equation}
\begin{split}
B_{\rm{tor}}(r,\theta,\phi)&= \sum_{lm}\frac{T(r)im}{r\sin\theta}c_{lm}P_{lm}e^{im\phi}\hat{\theta}\\
& -\sum_{lm}\frac{T(r)}{r}c_{lm}\dv{P_{lm}}{\theta}e^{im\phi}\hat{\phi}.
\label{eq:Btexp} 
\end{split}
\end{equation}
The sums $\sum_{lm}$ run from mode $l$ = 1 to $l = l_{\rm{max}}$, and from order $m = -l$ to $m = l$, where the maximum mode depends on the resolution of the data available. 

\subsection{Vector potential fields}
\label{sec:Acalc}

Having determined a general magnetic field expression, next we require the corresponding vector potential field $\boldsymbol{A}$:
\begin{equation}
\boldsymbol{B}=\nabla\times \boldsymbol{A}.
\label{eq:A_def}
\end{equation}
Given Equations \ref{eq:Bporiginal} and \ref{eq:Btoriginal}, it follows that the poloidal and toroidal components of the vector potential field are
\begin{equation}
\boldsymbol{A}_{\rm{pol}}=\curl{[\Phi\hat{r}]},
\label{eq:Aporiginal}
\end{equation}
and
\begin{equation}
\boldsymbol{A}_{\rm{tor}}=\Psi\hat{r}.
\label{eq:Atoriginal}
\end{equation}
Substituting Equations \ref{eq:Phi} and \ref{eq:Psi} for $\Phi$ and $\Psi$ and expanding results in: 
\begin{equation}
\begin{split}
A_{\rm{pol}}&=\sum_{lm}\frac{im}{r\sin\theta}S(r)c_{lm}P_{lm}e^{im\phi}\hat{\theta}\\
&-\sum_{lm}\frac{1}{r}S(r)c_{lm}\dv{P_{lm}}{\theta}e^{im\phi}\hat{\phi},
\end{split}
\label{eq:Apexp}
\end{equation}
\begin{equation}
A_{\rm{tor}}=\sum_{lm}T(r)c_{lm}P_{lm}e^{im\phi}\hat{r}.
\label{eq:Atexp}
\end{equation}

\subsection{Calculating the helicity density}
\label{sec:Hcalc}

When expanding Equation \ref{eq:hpoltorfull} we find that both $A_{\rm{pol}}\cdot B_{\rm{pol}}$ and $A_{\rm{tor}}\cdot B_{\rm{tor}}$ are zero. Consequently the helicity density equation simplifies to:  
\begin{equation}
h=\boldsymbol{A}_{\rm{pol}}\cdot \boldsymbol{B}_{\rm{tor}}+ \boldsymbol{A}_{\rm{tor}}\cdot \boldsymbol{B}_{\rm{pol}}. 
\label{eq:hpoltor} 
\end{equation}
By inserting the magnetic field from Section \ref{sec:Bpoltor} (Equations \ref{eq:Bpexp} and \ref{eq:Btexp}) and the vector potential from Section \ref{sec:Acalc} (Equations \ref{eq:Apexp} and \ref{eq:Atexp}), the real part of the magnetic helicity density is given by\footnote{The given expression for magnetic helicity density is derived using a left-handed coordinate system, it can be converted to a right-handed coordinate system simply by flipping the sign.}: 
\begin{equation}
\begin{split}
    h(r,\theta,\phi)=&\Re\bigg(\sum_{lm}\sum_{l^\prime m^\prime}\frac{1}{r^2}S(r)T(r)c_{lm}c_{l^\prime m^\prime}e^{i\phi(m+m^\prime)}\\
    &\bigg(P_{lm}P_{l^\prime m^\prime}\bigg(l(l+1)-\frac{mm^\prime}{\sin^2\theta}\bigg)+\dv{P_{lm}}{\theta}\dv{P_{l^\prime m^\prime}}{\theta}\bigg)\bigg).
\end{split}
\label{eq:generalh}
\end{equation}
As the helicity density is calculated by taking the dot product of sums running between the same limits, the toroidal components have been denoted with a prime in order to distinguish the different sums.  

The radial functions $S(r)$ and $T(r)$ are the only terms in Equation \ref{eq:generalh} which require information about the specific star considered. Using the ZDI technique \citep{Semel1989}, large scale magnetic fields can be determined at stellar surfaces (r = $R_\star$), which provides values for $S(R_\star)$ and $T(R_\star)$. Unfortunately it is unknown how the stellar magnetic field extends beyond this, hence the magnetic helicity density can only be evaluated at the stellar surface.  

The surface fields are decomposed into poloidal and toroidal components (e.g. \cite{Donati2006,Vidotto2016}):
\begin{equation}
\begin{split}
B_{\rm{pol}}(\theta,\phi)&=\sum_{lm}\alpha_{lm}c_{lm}P_{lm}e^{im\phi}\hat{r}\\
& +\sum_{lm}\frac{\beta_{lm}}{(l+1)}c_{lm}\dv{P_{lm}}{\theta}e^{im\phi}\hat{\theta}\\
&+\sum_{lm}\frac{\beta_{lm}im}{(l+1)\sin\theta}c_{lm}P_{lm}e^{im\phi}\hat{\phi},
\label{eq:sBp} 
\end{split}
\end{equation}
\begin{equation}
\begin{split}
B_{\rm{tor}}(\theta,\phi)&=\sum_{lm}\frac{\gamma_{lm}im}{(l+1)\sin\theta}c_{lm}P_{lm}e^{im\phi}\hat{\theta}\\
&-\sum_{lm}\frac{\gamma_{lm}}{(l+1)}c_{lm}\dv{P_{lm}}{\theta}e^{im\phi}\hat{\phi},
\label{eq:sBt} 
\end{split}
\end{equation}
which are characterised by $\alpha_{lm}$, $\beta_{lm}$, and $\gamma_{lm}$ coefficients. A detailed description of how these coefficients are determined given radial, meridional and azimuthal magnetic field components can be found in \cite{Vidotto2016}. Equating this surface field with the full magnetic field (Equations \ref{eq:Bpexp} and \ref{eq:Btexp}) evaluated at $r=R_\star$; $\boldsymbol{B}(\theta,\phi)=\boldsymbol{B}(R_\star,\theta,\phi)$, gives:
\begin{equation}
S(R_\star)=\frac{\alpha_{lm}R_\star^2}{l(l+1)},
\label{eq:S_r} 
\end{equation}
\begin{equation}
T(R_\star)=\frac{\gamma_{lm}R_\star}{(l+1)}.
\label{eq:T_r} 
\end{equation}

Having established expressions for $S(R_\star)$ and $T(R_\star)$, Equation \ref{eq:generalh} can be evaluated at $r=R_\star$, which gives the magnetic helicity density at any point ($\theta$,$\phi$) on the stellar surface: 
\begin{equation}
\begin{split}
    h(R_\star,\theta,\phi)&=\Re\bigg(\sum_{lm}\sum_{l^\prime m^\prime}\frac{\alpha_{lm}\gamma_{l^\prime m^\prime}R_\star}{(l^\prime+1)l(l+1)}c_{lm}c_{l^\prime m^\prime}e^{i\phi(m+m^\prime)}\\
    &\bigg(P_{lm}P_{l^\prime m^\prime}\bigg(l(l+1)-\frac{mm^\prime}{\sin^2\theta}\bigg)+\dv{P_{lm}}{\theta}\dv{P_{l^\prime m^\prime}}{\theta}\bigg)\bigg).
\end{split}
\label{eq:surfaceh}
\end{equation}
%
This expression for helicity density can be applied to any star given only its stellar radius and the $\alpha_{lm}$ and $\gamma_{lm}$ coefficients characterising its poloidal and toroidal magnetic field components.  

We note that $\beta_{lm}$ does not appear in the helicity density equation. This is due to helicity being the linking of toroidal and poloidal fields. The toroidal field lines lie on spherical surfaces, and the poloidal field lines pass through these surfaces. However, only the radial component of the poloidal field links through the toroidal field, the $\theta$ and $\phi$ components lie on the same spherical surfaces as the toroidal field and so provide no ``linkage''. Since the radial part of the poloidal field depends only on $\alpha_{lm}$, and the toroidal field depends only on $\gamma_{lm}$, $\beta_{lm}$ is not needed.

When comparing the helicity density of different stars, or at different times, rather than considering the helicity density at specific points $(\theta,\phi)$ on the stellar surface, it is more convenient to calculate an average helicity density value across some surface area A: 
\begin{equation}
\langle{h}\rangle= \frac{\int h(R_\star,\theta,\phi)\,\mathrm{d}A}{A}. 
\label{eq:av_h} 
\end{equation}

\section{Application}
\label{sec:Application}

\subsection{Observational and simulated data}
\begin{center}
\begin{table*}
\centering
 \caption{Our stellar sample. From left to right the columns show: star name, mass, radius, rotation period, Rossby number, absolute helicity density calculated up to $l\leq 4$ averaged across the visible hemisphere (a mean value if more than one magnetic map is available), $l_{\rm{max}}$, number of magnetic maps used and references to those magnetic maps. A more comprehensive table can be found in \citet{Vidotto2014}.}
\begin{tabularx}{\textwidth}{ l X X X X X X X X r}
    \hline
        Star ID & $M_{\star}$  & $R_{\star}$ & $P_{\rm{rot}}$ & R$_{o}$&Age&$|\langle{h\,}\rangle|_{l\leq4}$& $l_{\rm{max}}$ & No. of & Ref.\T\\ 
    & (M$_{\odot}$) & (R$_{\odot}$) & (d) &&(Myr)&(Mx$^2\rm{cm}^{-3}$)  & &maps &\B \\
    \hline
    \multicolumn{10}{l}{\textbf{Solar-like stars}\T} \\
HD 3651	&	0.88	&	0.88	&	43.4& 1.916	&8200& 1.938e+11 &10 & 1 & 1	\\
HD 9986	&	1.02	&	1.04	&	23.0&1.621	&4300&7.552e+09& 10 & 1 & 1	\\
HD 10476	&	0.82	&	0.82	&	16.0&0.576	&8700&8.060e+09& 10 &1 & 1	\\
HD 20630	&	1.03	&	0.95	&	9.30&0.593	&600&3.108e+13& 10 &1& 2	\\
HD 22049	&	0.86	&	0.77	&	10.3&0.366	&440&8.056e+11& 10 &1& 3	\\ 
HD 39587	&	1.03	&	1.05	&	4.83&0.295	&500&2.379e+12& 10 &1& 1	\\ 
HD 56124	&	1.03	&	1.01	&	18.0&	1.307	&4500&2.889e+11& 10 &1& 1	\\ 
HD 72905	&	1	&	1	&	5.00&	0.272	&500&1.634e+13& 10 &1& 1	\\
HD 73350	&	1.04	&	0.98	&	12.3&	0.777	&510&3.771e+11& 10 &1& 1	\\
HD 75332	&	1.21	&	1.24	&	4.80	&	>1.105	&1800&1.230e+12& 15 &1& 1	\\
HD 78366	&	1.34	&	1.03	&	11.4&	>2.781	&2500&1.305e+12& 10 &3& 4	\\
HD 101501	&	0.85	&	0.9	&	17.6&	0.663	&5100&3.267e+12& 10 &1& 1	\\
HD 131156A	&	0.93	&	0.84	&	5.56&	0.256	&2000&2.539e+13& 10 &7& 5, 6	\\

HD 131156B	&	0.99	&	1.07	&	10.3&	0.611	&2000&1.302e+13& 10 &1& 1	\\
HD 146233	&	0.98	&	1.02	&	22.7&	1.324	&4700&9.357e+08& 10 &1& 7	\\ 
HD 166435	&	1.04	&	0.99	&	3.43&	0.259	&3800&5.368e+12& 10 &1& 1	\\
HD 175726	&	1.06	&	1.06	&	3.92&	0.272	&500&2.139e+12& 10 &1& 1	\\
HD 190771	&	0.96	&	0.98	&	8.80&	0.453	&2700&5.472e+12& 10 &1& 7	\\
HD 201091A	&	0.66	&	0.62	&	34.2&	0.786	&3600&9.288e+10& 10 &1& 8	\\ 
HD 206860	&	1.10	&	1.04	&	4.55&	0.388	&260&3.108e+13& 10 &1& 9	\\
\multicolumn{10}{l}{\textbf{Young Suns}} \\
AB Dor 	&	0.76	&	0.96	&	0.5&	0.026	&120&2.318e+14& 25 &6& 10	\\ 
BD-16351	&	0.9	&	0.83	&	3.39&	-	&30&6.257e+13& 15 &1& 11	\\
HII 296	&	0.8	&	0.74	&	2.61&	-	&130&3.731e+13& 15 &1& 11	\\
HII 739	&	1.08	&	1.03	&	2.7&	-	&130&1.249e+12& 15 &1& 11	\\
HIP 12545	&	0.58	&	0.57	&	4.83&	-	&21&5.216e+14& 15 &1& 11	\\
HIP 76768	&	0.61	&	0.6	&	3.64&	-	&120&4.058e+14& 15 &1& 11	\\
TYC 0486-4943-1	&	0.69	&	0.68	&	3.75&	-	&120&4.847e+12& 15 &1& 11	\\
TYC 5164-567-1	&	0.85	&	0.79	&	4.71&	-	&120&4.150e+13& 15 &1& 11	\\
TYC 6349-0200-1	&	0.54	&	0.54	&	3.39&	-	&21&2.305e+13& 15 &1& 11	\\
TYC 6878-0195-1	&	0.65	&	0.64	&	5.72&	-	&21&5.125e+13& 15 &1& 11	\\
\multicolumn{10}{l}{\textbf{Hot Jupiter Hosts}} \\
$\tau$ Boo	&	1.34	&	1.42	&	3&	>0.732	&2500&1.722e+11&8,5&6& 12, 13, 14, 15	\\
HD 73256	&	1.05	&	0.89	&	14&	0.962	&830&3.619e+11& 4 &1& 15	\\
HD 102195	&	0.87	&	0.82	&	12.3&	0.473	&2400&2.687e+12& 4 &1& 15	\\
HD 130322	&	0.79	&	0.83	&	26.1&	0.782	&930&1.277e+11& 4 &1& 15	\\
HD 179949	&	1.21	&	1.19	&	7.6&	>1.726	&2100&6.133e+10& 6 &1& 16	\\
HD 189733 	&	0.82	&	0.76	&	12.5&	0.403	&600&8.669e+12& 5 &2& 17	\\
\multicolumn{10}{l}{\textbf{M dwarf Stars}} \\													
GJ 569A	&	0.48	&	0.43	&	14.7&	<0.288	&130&1.807e+14& 5 &1& 18	\\
GJ 410	&	0.58	&	0.52	&	14&	<0.267	&710&1.532e+14& 5 &2& 18	\\
GJ 182	&	0.75	&	0.82	&	4.35&	0.054	&21&5.905e+14& 8 &1& 18	\\
GJ 49	&	0.57	&	0.51	&	18.6&	<0.352	&1200&2.494e+13& 5 &1& 18	\\
GJ 494A	&	0.59	&	0.53	&	2.85&	0.092	&-&1.644e+14& 8 &2& 18	\\
GJ 388	&	0.42	&	0.38	&	2.24&	0.047	&-&1.238e+14& 8 &2& 19	\\
EQ Peg A	&	0.39	&	0.35	&	1.06&	0.02	&-&4.550e+14& 4 &1& 19	\\
EQ Peg B	&	0.25	&	0.25	&	0.4&	0.005	&-&4.724e+14& 8 &1& 19	\\
GJ 873	&	0.32	&	0.3	&	4.37&	0.068	&-&8.771e+14& 8 &2& 19	\\
GJ 9520 	&	0.55	&	0.49	&	3.4&	0.097	&-&2.173e+14& 8 &2& 18	\\
V374 Peg &	0.28	&	0.28	&	0.45&	0.006	&-&9.861e+13& 10 &2& 20	\\
GJ 1111 & 0.1 & 0.11 & 0.46& 0.0059 &-&1.533e+13& 6 &3& 21 \\ 
GJ 1156 & 0.14 & 0.16 & 0.49 & 0.0081 &-&1.076e+13& 6 &3& 21 \\
GJ 1245 B & 0.12 & 0.14 & 0.71& 0.011 &-&1.565e+13& 4 &2& 21 \\
WX UMa & 0.1 & 0.12 & 0.78& 0.01 &-&1.599e+15& 4 &2& 21 \\

\hline
\end{tabularx}
\begin{flushleft} 
1: Petit et al. (in prep); 2: \citet{DoNascimento2016}; 3: \citet{Jeffers2014}; 4: \citet{Morgenthaler2011}; 5: \citet{Morgenthaler2012}; 6: Jeffers et al. (in prep); 7: \citet{Petit2008}; 8: \citet{Boro2016}; 9: \citet{Boro2015}; 10: \citet{Donati2003}; 11: \citet{Folsom2016}; 12: \citet{Catala2007}; 13: \citet{Donati2008b}; 14: \citet{Fares2009}; 15: \citet{Fares2013}; 16: \citet{Fares2012}; 17: \citet{Fares2010}; 18: \citet{Donati2008}; 19: \citet{Morin2008b}; 20: \citet{Morin2008}; 21: \citet{Morin2010}  
\end{flushleft}
  \label{table:StellarSample}
\end{table*}
\end{center}
In this paper we calculate the large-scale helicity density of the Sun, as well as 51 additional stars listed in Table \ref{table:StellarSample}. The magnetic maps we use for each star are referenced in the last column of the table. From left to right the remaining columns show the name of the star, stellar mass, stellar radius, rotation period, Rossby number, age, absolute helicity density ($l\leq 4$) averaged across the visible stellar hemisphere, the maximum $l$-mode and the number of magnetic maps used. For references to the stellar parameters listed, as well as a more detailed table with further information on these stars see \cite{Vidotto2014}. The stellar sample consists of stars with spectral types F, G, K and M, with masses ranging from 0.1-1.34 M$_\odot$. The resolutions of the magnetic maps of the stellar sample range from $l_{\rm{max}}$ = 4 to 25, with higher modes indicating a higher resolution, meaning smaller scale magnetic fields are detected. The stars with $l_{\rm{max}}$ < 8, which comprise the majority of the M-dwarfs and hot Jupiter hosts in Table \ref{table:StellarSample}, will be omitted for part of the analysis in Section \ref{sec:stellarH} where we choose to consider stars with $l_{\rm{max}}$ = 8 to allow for a larger range of $l$-modes.  

In order to place the Sun in context, we include it in our study. The solar magnetic maps we use come from observations taken by the Helioseismic and Magnetic Imager (HMI) on-board the Solar Dynamics Observatory (SDO) \citep{Scherrer2012,Pesnell2012}. We also use surface magnetograms taken from the 3D non-potential magnetic field simulation presented in \cite{Yeates2012}. Together the data spans almost two whole solar cycles; the observed solar data covers most of solar cycle 24, and the simulation is over solar cycle 23. There is a slight overlap in time between the two data sets, which proves useful when checking for consistency between the simulated and observed data. The ability to follow variations in helicity density through a cycle may provide insights into the sources of the scatter in stellar values.

Throughout Section \ref{sec:solarH} the solar helicity will be calculated up to $l$ = 8 only, as this is a reasonable resolution for most of the other stars in our sample. Both the observed and simulated Sun can be recovered to a higher $l_{\rm{max}}$, but this large-scale helicity density is a reasonable approximation of what we would detect if we could observe the Sun as a star. Only the helicity captured on the largest scales is shown - the contribution from smaller-scale features such as active regions is omitted. For the sake of consistency, we compare the stellar and solar data using the same number of $l$-modes. The lowest resolution used will be $l\leq$ 4, as this is the highest common mode of the stellar sample in Table \ref{table:StellarSample}. 

\subsection{Large-scale solar helicity densities}
\label{sec:solarH}

\begin{figure*}\centering\hspace*{\fill}
	\includegraphics[width=\textwidth]{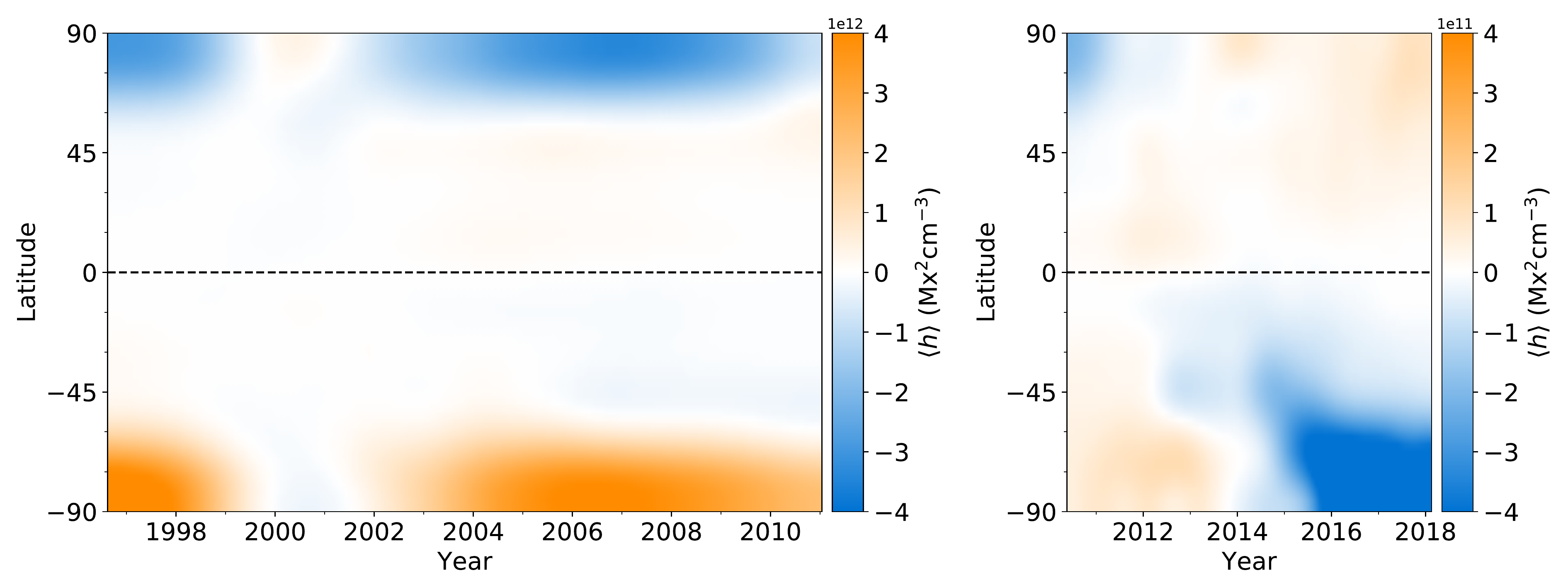}\hspace*{\fill}
	\caption{The evolution in time of the average solar helicity density for $l\leq$ 8 at each latitude, using simulated data (\textit{left}) and observational data (\textit{right}). Gaussian smoothing has been applied to remove small variations and highlight overall trends. The colour tables saturate at $\pm4\times10^{12}$ Mx$^2$cm$^{-3}$ (\textit{left}) and $\pm4\times10^{11}$ Mx$^2$cm$^{-3}$ (\textit{right}).}
	\label{fig:h_vs_t}
\end{figure*}

We calculate the longitudinally averaged helicity density for every observed and simulated magnetic map and plot it as a function of time in Fig. \ref{fig:h_vs_t}. We use Gaussian smoothing to remove small variations and highlight the overall trends. The helicity densities in the plots are limited to $\pm4\times10^{12}$ Mx$^2$cm$^{-3}$ (left panel) and $\pm4\times10^{11}$ Mx$^2$cm$^{-3}$ (right panel) which results in saturation, but reveals more structure towards the equator. The resulting pattern is consistent with the large scale magnetic helicity density presented by \cite{Pipin2019} in their paper exploring the evolution of the solar magnetic helicity density throughout solar cycle 24 (see their Fig. 5a). Our plot has the opposite polarity to theirs, which may be due to differing sign conventions; using a different coordinate system can change the sign of the helicity density. 

The helicity density at the poles of the simulated Sun is approximately a full order of magnitude stronger than that of the observed Sun. However, apart from the magnitude, the two plots show consistent results and match up nicely in the overlapping year ($\sim$ 2010-2011) showing a positive south pole and a negative north pole. In both cases, the strong signal at the poles overshadows most structure around the equator and the sign of the helicity density flips across the equator. The helicity is predominantly negative in the northern hemisphere and positive in the southern hemisphere until $\sim$ 2014, when it reverses. A much shorter sign reversal can also be seen around 2000. It is perhaps worthwhile to note that 2000 and 2014 are the years with the highest sunspot activity during solar cycles 23 and 24 respectively\footnote{http://www.sidc.be/silso/datafiles - Database of the monthly mean total sunspot number (1749 - now), accessed on 12 Dec 2019.}, we cannot confidently state whether this is coincidence or consequence.  

\begin{figure*}\centering\hspace*{\fill}
	\includegraphics[width=\textwidth]{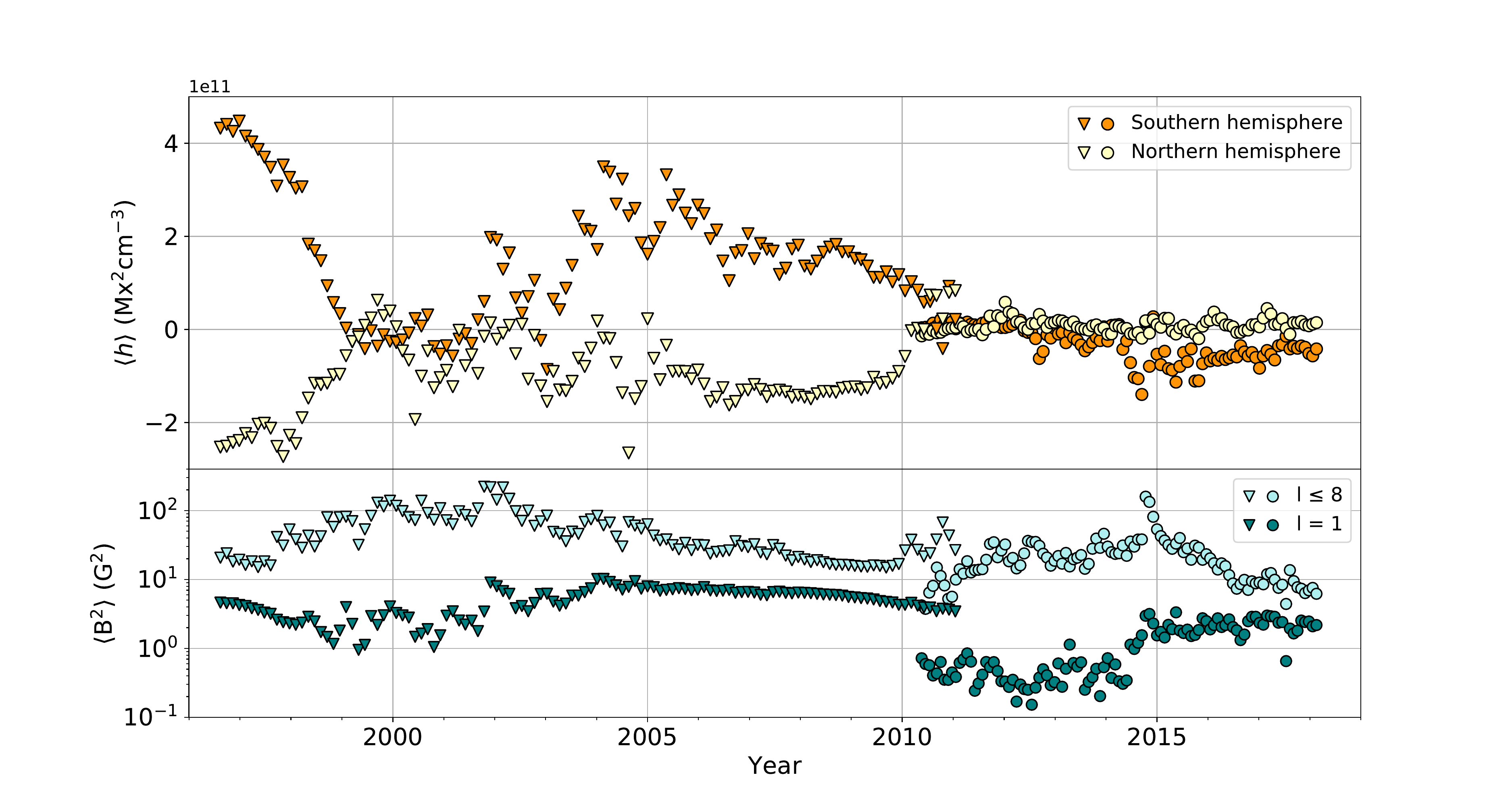}\hspace*{\fill}
	\caption{\textit{Top}: The evolution in time of the average solar helicity density for $l\leq$ 8 across the southern (orange) and northern (yellow) hemispheres \textit{Bottom}: The mean squared magnetic flux density for $l\leq$ 8 (light teal) and $l$ = 1 (dark teal) across the same time period. The triangles and circles correspond to results based on simulated and observational data respectively.}
	\label{fig:Hsolarcycle}
\end{figure*}

Calculating the average helicity density across both hemispheres allows us to compare the two more quantitatively. The top panel of Fig. \ref{fig:Hsolarcycle} shows the helicity density over time averaged across the southern (orange) and northern (yellow) hemispheres respectively. The triangles correspond to the simulated Sun, and the circles correspond to the HMI observations. Despite having already established that the helicity density flips signs across the equator, this plot reveals more clearly that the overall helicity density of the Sun is never exactly zero. In the case of the simulated Sun, the helicity density is approximately mirrored across the equator, and over time it will average to more or less zero. For the observed Sun on the other hand the helicity density in the southern hemisphere dominates throughout the second half of the time period. This could be due to computational errors, or there could be a real imbalance. \cite{Yang2012} showed an asymmetry between the large-scale magnetic helicity fluxes in the northern and southern hemisphere across solar cycle 23, which would lead to different amounts of helicity accumulating in each hemisphere. We propose, due to the sign change across the equator, the helicity density averaged across a single hemisphere provides a more meaningful result than an average across the entire sphere, which represents a residual value.

The bottom panel of Fig. \ref{fig:Hsolarcycle} shows how the solar magnetic energy for $l$ = 1 and $l\leq$ 8 behaves throughout the same time period over which we are presenting the helicity density in the top panel. The mean squared magnetic flux density across the Sun, $\langle{\rm{B}^2_{}\,}\rangle$, represents a proxy for magnetic energy. The energies of both the simulation and the real Sun follow a similar pattern. On the approach to cycle maximum, the total energy (summed up to mode $l = 8$) increases, while the dipole energy decreases. In the declining phases, as the total energy decreases, the dipole energy first grows, then decreases slowly to reach another minimum at cycle maximum. The helicity density also follows a cyclic pattern, although shifted later than the total energy by about one quarter in phase. 

\subsection{Large scale stellar helicity densities}
\label{sec:stellarH}

Based on the exploration of solar helicity density in the previous section we choose to compare the helicity densities of our stellar sample in terms of their averages across a single hemisphere. We choose the hemisphere pointing towards the observer, noting that the other is partially obscured, and even partially invisible. Because the sign of the helicity can change across hemispheres or over time, once we have calculated the average we take its absolute value. In the case of the observed solar data we select the southern hemisphere. When the Sun is plotted for comparison throughout this section, we are using the observations of the real Sun between $\sim$ 2010-2018, not the simulated data. 

\begin{figure}\centering\hspace*{\fill}
	\includegraphics[width=\columnwidth]{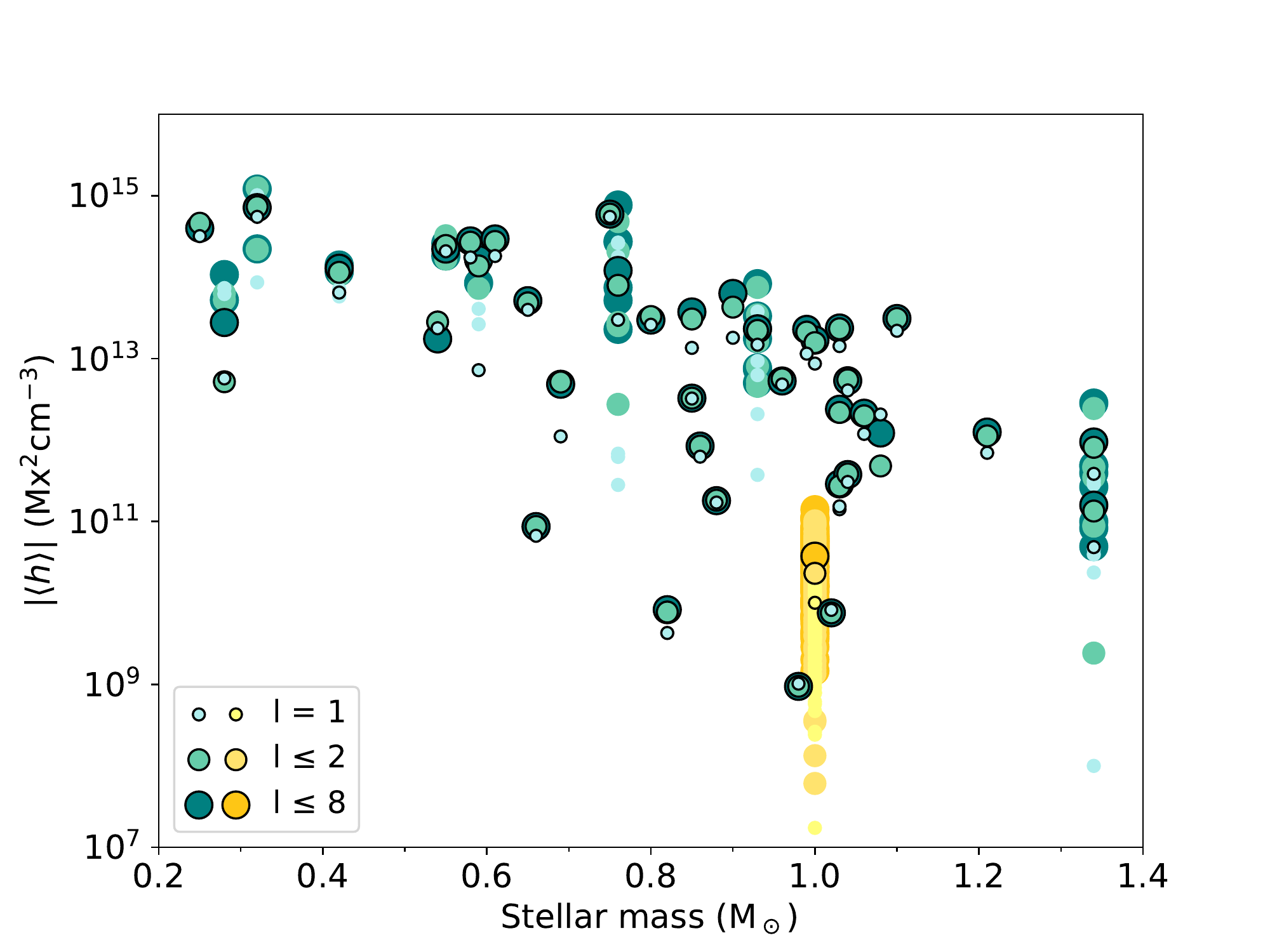}\hspace*{\fill}
	\caption{Absolute helicity densities averaged across a single hemisphere as a function of stellar mass. The orange shades show the Sun across the southern hemisphere between $\sim$ 2010-2018, and the teal shades show the remaining stars in the stellar sample. The helicity densities are calculated up to different modes, which are represented by different sizes and colours. $l_{\rm{max}}$ increases from small and light to large and dark. Symbols without an outline represent multiple measurements for the same stars, and the symbols with black edges are average values.}
	\label{fig:hmodes}
\end{figure}

To explore the importance of the chosen resolution, we plot the absolute average helicity density calculated up to three different $l$-mode limits as a function of stellar mass in Fig. \ref{fig:hmodes}. The different colours correspond to $l$ = 1 (dipole), $l\leq$ 2 (quadrupole) and $l\leq$ 8. Consequently the stars in Table \ref{table:StellarSample} with $l_{\rm{max}}$ < 8, are excluded from this plot. The points without black outlines show multiple values for the same star, and the points with black outlines represent average values. An example of this is the Sun, shown in orange shades, where the points with outlines are the average values across $\sim$ 2010-2018. We find that the helicity densities recovered using dipole or quadrupole fields alone result in good representations of the $l\leq$ 8 helicity density. On average the $l$ = 1 and $l\leq$ 2 points deviate by $\sim$ 40 and 13 \% from the $l\leq$ 8 points. Including higher-order modes changes the magnitude of the helicity density only slightly, and the general trend across stellar masses remains the same. Hence, for the remainder of this paper we limit ourselves to $l\leq$ 4 in order to increase the number of eligible stars and particularly capture the hot Jupiter hosts and M dwarfs better. 

From Fig. \ref{fig:hmodes} we find the absolute helicity density increases with decreasing stellar mass, reaching a plateau for stars of M$_\star \lesssim$ 0.5 M$_\odot$. In fact, a number of magnetic field properties have been discovered to change across this 0.5 M$_\odot$ boundary \citep{Donati2008,Morin2008b,Morin2010,See2015}. The authors suggest it is related to the onset of the sharp transition in internal stellar structure from partially convective stars with inner radiative interiors out to $\sim$ 0.5 R$_\star$ (at 0.5 M$_\odot$) to fully convective stars (at 0.3 M$_\odot$). As noted by \citet{See2015}, there is a correlation between stellar mass and rotation period for our ZDI sample. Generally the more massive stars in our sample tend to be spinning slower (there are exceptions; we do have some high-mass stars that rotate very fast e.g. AB Dor). This makes it hard to say whether the helicity trend is with mass or rotation. In an attempt to shed light on this, we consider the Rossby number, R$_o$; a parameter which encapsulates information about both mass and rotation period.

\begin{figure}\centering\hspace*{\fill}
	\includegraphics[width=\columnwidth]{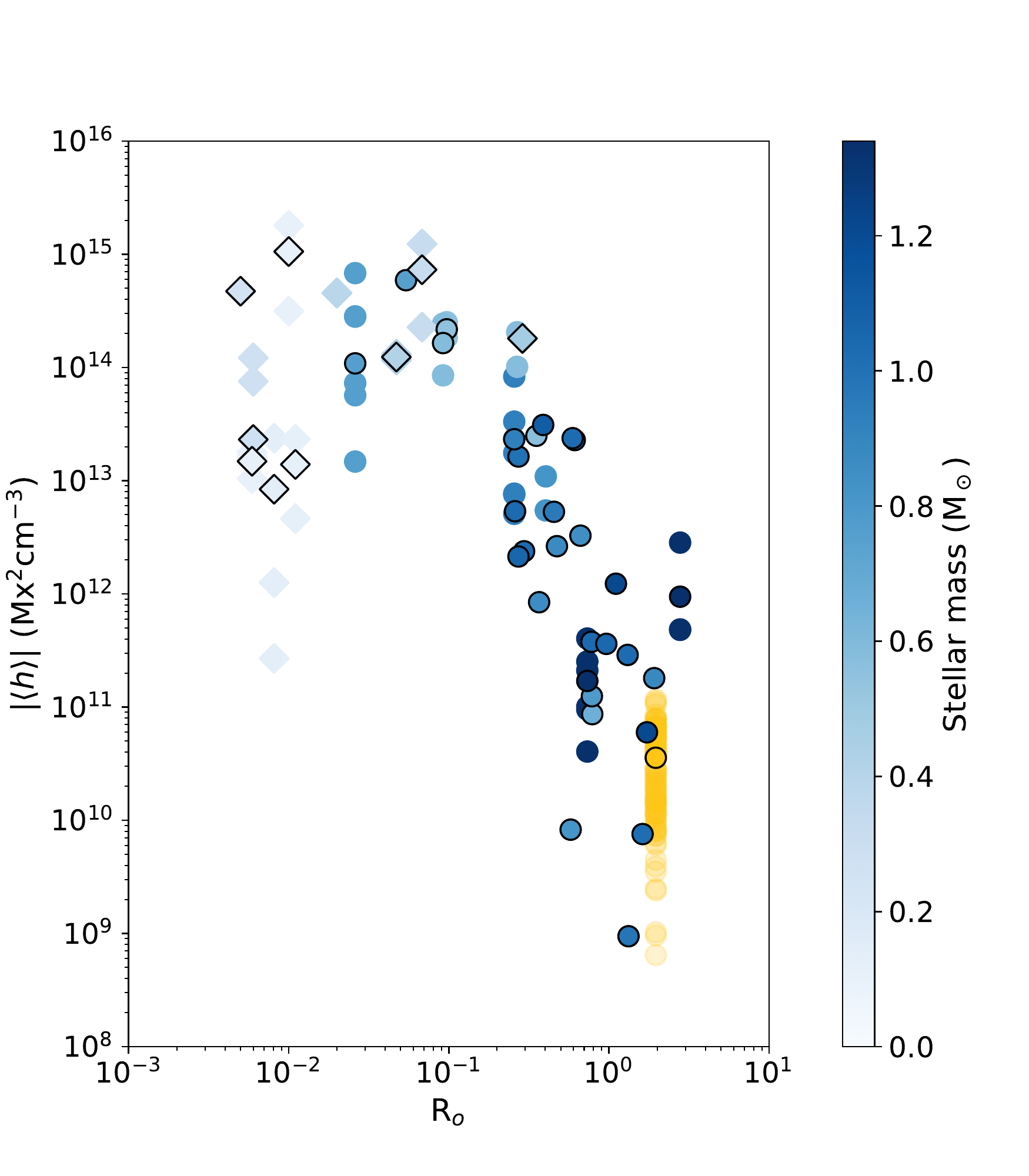}\hspace*{\fill}
	\caption{Absolute helicity density ($l\leq$ 4) averaged across a single hemisphere versus Rossby number. The colour of the symbols correspond to stellar mass, and the diamonds represent stars with $M_\star$ < 0.5 $M_\odot$. Symbols without an outline represent multiple measurements for the same stars, and the symbols with black edges are average values. The orange circles show the range of solar values, for the southern hemisphere, between $\sim$ 2010-2018.}
	\label{fig:HRossby}
\end{figure}

Fig. \ref{fig:HRossby} shows the absolute average helicity density
versus Rossby number. We use the Rossby numbers listed in Table \ref{table:StellarSample}, originally calculated and published in \cite{Vidotto2014}. Stars for which we lack Rossby number estimates are omitted from this plot. Stellar mass is denoted by the colour of the symbols. The sample of stars is divided at 0.5 M$_\odot$; circles represent stars with masses higher than 0.5 M$_\odot$, and diamonds represent stars with lower masses. The orange circles show the range of solar values, assuming a solar Rossby number of 1.96 \citep{Cranmer2011}. As in Fig. \ref{fig:hmodes}, the black outlines indicate average values.  The helicity density increases as the Rossby number decreases until a maximum is reached around $\rm{R}_o\sim 0.1$. Whether this is a true maximum, or simply a saturation is not clear. The same behaviour and saturation point has been reported for other stellar properties such as the ratio of X-ray to bolometric luminosities \citep{Wright2011} and the toroidal and poloidal magnetic energy densities \citep{See2015}. The stars with masses below 0.5 M$_\odot$ are all grouped together in the saturated region, alongside one higher mass, rapidly rotating star at R$_o$ = 0.026 (AB Dor). 

\begin{figure*}
\centering\hspace*{\fill}

	\includegraphics[width=\columnwidth]{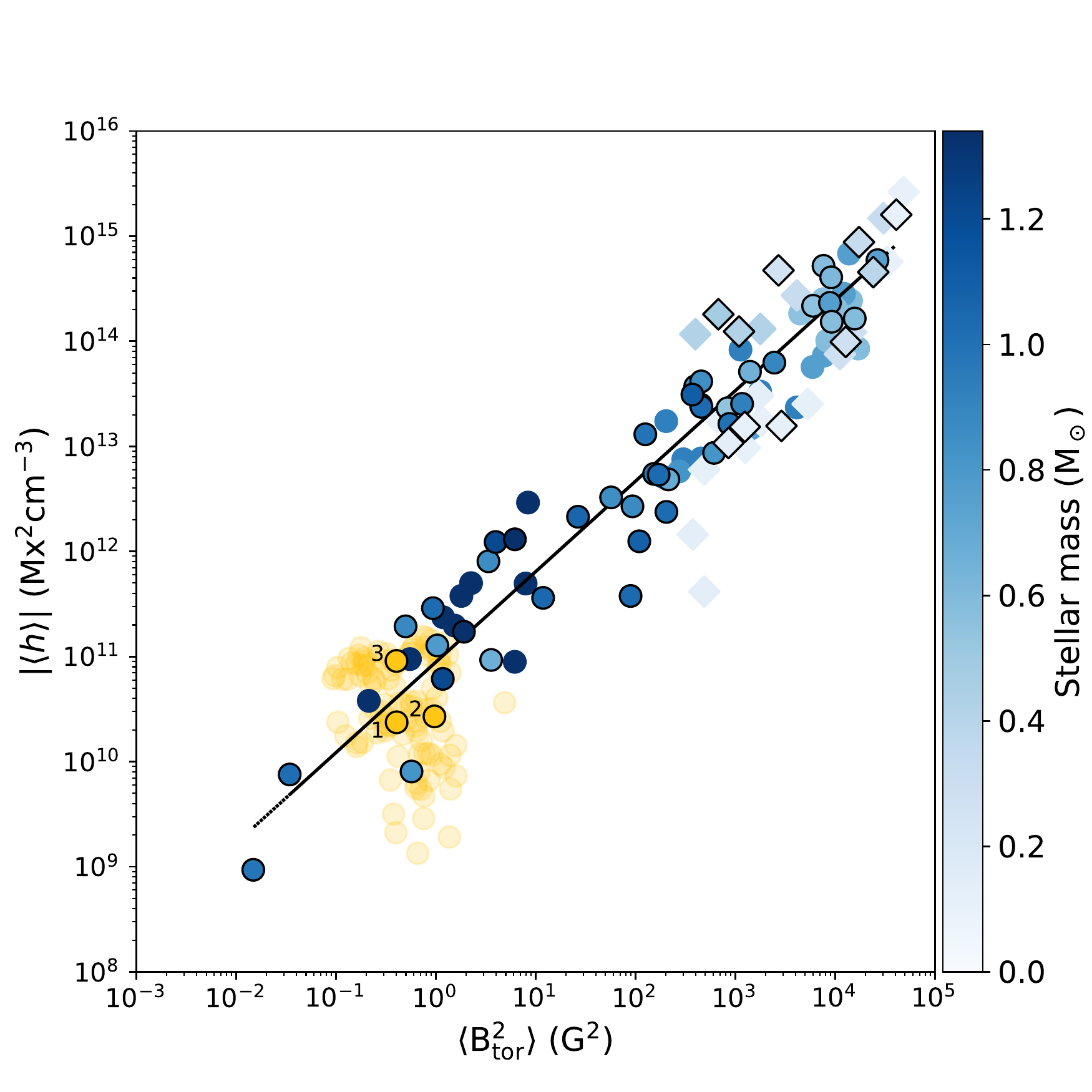}\hspace*{\fill}
		\includegraphics[width=\columnwidth]{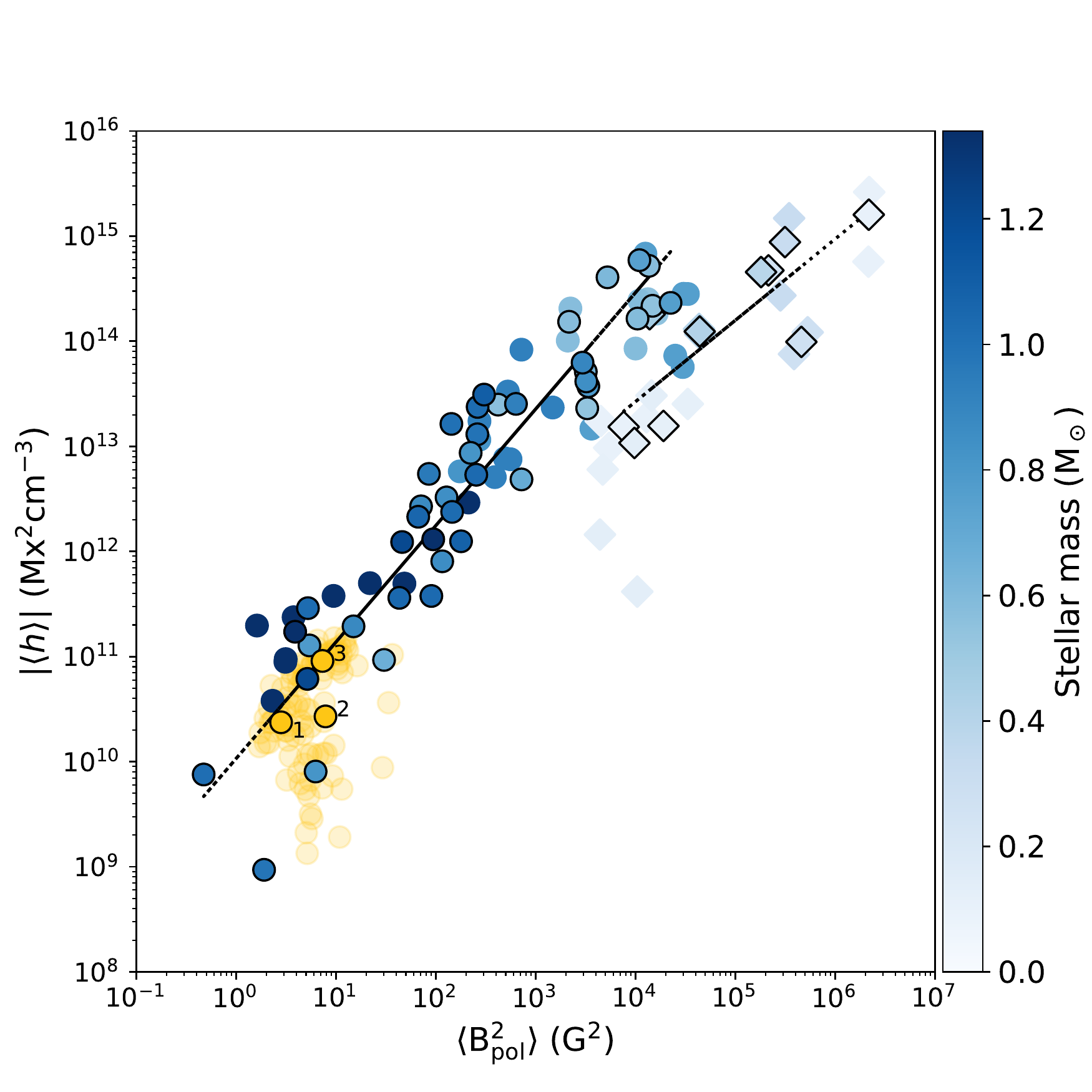}\hspace*{\fill}
	
	\caption{Absolute helicity density for $l\leq$ 4 averaged across a single hemisphere versus the mean squared toroidal (\textit{left}) and poloidal (\textit{right}) magnetic flux densities across the star. The symbols are the same as in Fig. \ref{fig:HRossby}. Mean values of the solar data are given for the periods $\sim$ 2010-2012, 2012-2015 and 2015-2018, labelled 1, 2 and 3 respectively. The dashed lines show the best fit of |$\langle{h\,}\rangle$| $\propto$ $\langle{B^2\,}\rangle^\alpha$ calculated using the average values only. In the toroidal case one fit is given for all the stars, which results in $\alpha$ = 0.86 $\pm$ 0.04. In the poloidal case using all stars would produce a similar value of $\alpha$ = 0.85 $\pm$ 0.06, but every star with $M_\star$ < 0.5 $M_\odot$ would fall below the line and the overall fit would not be as tight. Consequently best fit lines of stars with $M_\star$ < 0.5 $M_\odot$ and $M_\star$ > 0.5 $M_\odot$ are shown separately, resulting in $\alpha$ = 0.77 $\pm$ 0.18 and $\alpha$ = 1.11 $\pm$ 0.07 respectively.}
	\label{fig:HvsB2}
\end{figure*}

\citet{See2015} found that the toroidal magnetic energy density increases faster with Rossby number than the poloidal magnetic energy density. We therefore plot the absolute average helicity density separately against toroidal and poloidal magnetic energy, see Fig. \ref{fig:HvsB2}. Again the mean squared magnetic flux density across the star, $\langle{\rm{B}^2_{}\,}\rangle$, acts as a proxy for magnetic energy. The shades of blue correspond to stellar mass, and the diamonds indicate stars with M$_\star$ < 0.5 M$_\odot$. The symbols without an outline represent multiple measurements of the same star, and the symbols with a black outline are average values. Values for the Sun spanning $\sim$ 2010-2018 are shown in orange, with the circles labelled 1, 2 and 3 being mean values for the periods $\sim$ 2010-2012, 2013-2015 and 2015-2018, denoting approximately the rising, maximum and declining phases of the cycle. These points show broadly how the Sun's position on the plot evolves in time. It is notable that the solar cyclic variation is within the scatter in values for other stars. 

We find that when the helicity density is plotted against the toroidal magnetic energy, all stars follow the relation $|\langle{h\,}\rangle|$ $\propto$ $\langle{\rm{B}^2_{\rm{tor}}\,\rangle}^{0.86\,\pm\,0.04}$ regardless of their interior structure. In contrast, when helicity density is plotted against the poloidal magnetic energy, there is a much larger scatter, and the low mass (largely convective) stars appear to lie on a different slope to those that have a radiative interior. 

\begin{figure*}\centering\hspace*{\fill}
	\includegraphics[width=\textwidth]{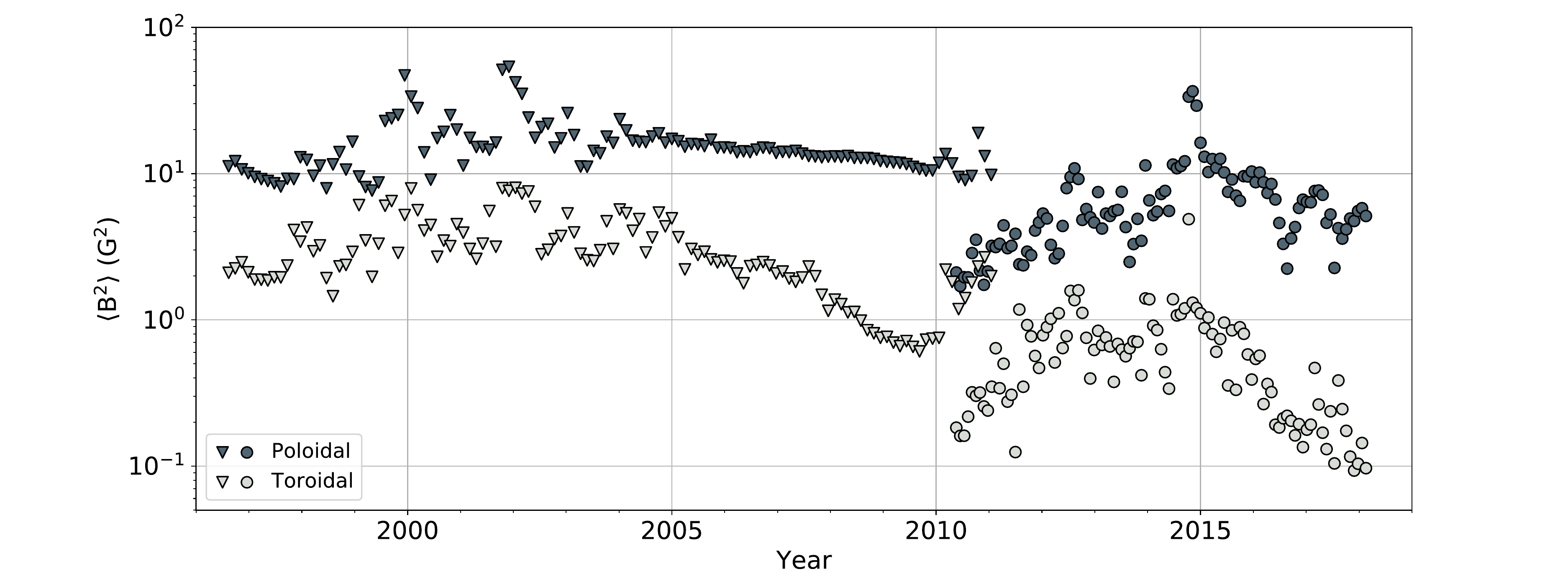}\hspace*{\fill}
	\caption{The mean squared magnetic flux density ($l\leq$ 4) of the poloidal (dark grey) and toroidal (light grey) magnetic field components through the solar cycle. The triangles and circles correspond to results based on simulated and observational data respectively.}
	\label{fig:Esolarcycle}
\end{figure*}

The evolution with time of the solar field appears to be different for the poloidal and toroidal components. The toroidal field moves from one side to the other of the best-fit line, whereas the poloidal field mainly stays to the right of the best-fit line. This can be understood by considering the variation of the large-scale poloidal and toroidal fields plotted against time in Fig. \ref{fig:Esolarcycle}. The two components are shown to vary together, but out of phase.
 
\begin{figure}\centering
	\includegraphics[width=\columnwidth]{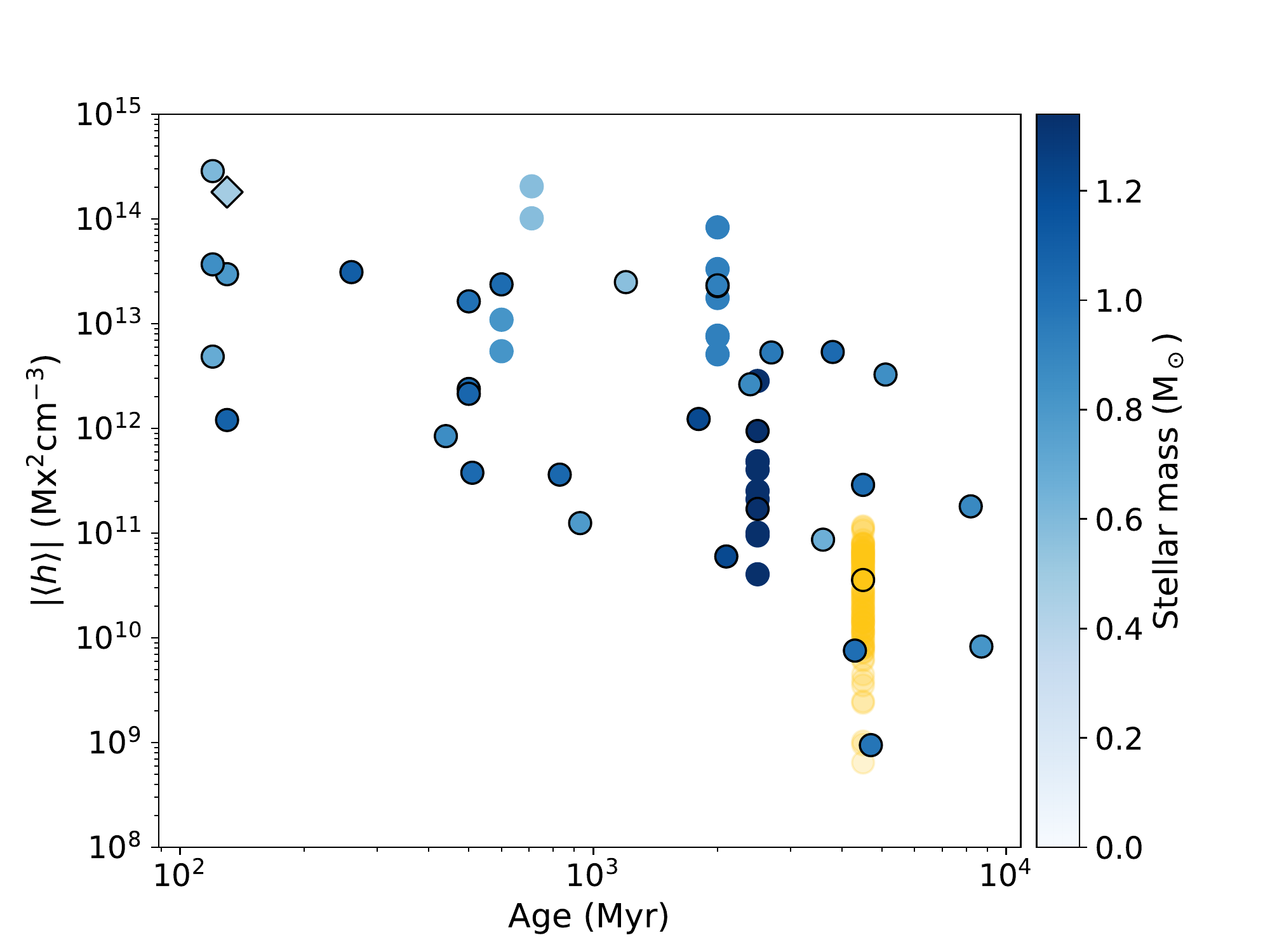}
	\caption{Absolute helicity density ($l\leq$ 4) averaged across a single hemisphere versus stellar age. Symbols are the same as in Figure \ref{fig:HRossby}.}
	\label{fig:hvsage}
\end{figure}

As stellar age is known to affect the magnetic properties of stars we plot the absolute average helicity density against age in Fig. \ref{fig:hvsage}. We use the ages listed in Table \ref{table:StellarSample}, more information and references for these can be found in \cite{Vidotto2014}. We lack ages for some of the M dwarfs in our sample, hence these are omitted from the plot. Despite a large spread of values, the Figure shows a clear decline in helicity density with age. Given the correlation outlined earlier between helicity density and magnetic field strength this result reflects the decline of field strength with age.

\section{Discussion}
\label{sec:Discussion}

By truncating our spherical harmonic expansion of the solar magnetic field at low $l$-values we have explored the variation in the solar helicity density that could be detected if the Sun were observed as a star. While this does not give us a complete picture of the Sun's helicity density, it allows us to determine the imprint left on the largest scales of the cyclic growth and decay in the solar helicity. This also provides a context within which to view the values we measure for other stars. 

Fig. \ref{fig:hmodes} demonstrates clearly the rise in helicity density with decreasing mass that would be expected in light of the high field strengths of many of the lowest mass stars \citep{Donati2008,Morin2008b,Morin2010,See2015}. At first glance the Sun appears to show an anomalously-low helicity density for its mass. Showing the variation with Rossby number instead (Fig. \ref{fig:HRossby}) clarifies this however. The Sun is simply a slower rotator than many other stars in our sample of similar mass. Its variation through its cycle is entirely within the scatter of the other stars. What is perhaps more intriguing about this Figure, however, is the possible existence of a peak in the helicity density at around R$_o \approx$ 0.1. This may of course be a plateau, rather than a peak. It is possible that the apparent maximum is a 
manifestation of bi-stability within the dynamo, leading to two possible regimes; one of weak field and one of strong field \citep{Morin2011}. Further observations are required to confirm this. Several other activity indicators also appear to peak in this parameter range. Super-saturation in X-ray emission \citep{James2000} has already been suggested for G and K dwarfs, although its existence in M dwarfs is not confirmed \citep{Jeffries2011,Wright2011} and more recently the possibility of a peak in the rate of large M-dwarf flares at R$_o\approx$ 0.1 has been suggested \citep{Mondrik2019}. This is also the regime in which \citet{See2017} find the maximum mass and angular momentum loss rates. This may mark a transition in the geometry of the magnetic fields with some of the lowest Rossby number stars lying in the ``bi-stable'' regime \citep{Morin2011,Schrinner2012,Gastine2013} where stars with similar parameters may exhibit either strong, simple fields, or weaker, more complex ones. A cyclic variation between these two states has also been proposed \citep{Kitchatinov2014}. 

Fig. \ref{fig:HvsB2}  shows a very tight correlation between the large-scale helicity density and toroidal energy density. Since helicity measures the linkage of the poloidal and toroidal fields, this suggests that a common process governs the field geometry of stars in the mass range $0.1-1.34$ M$_\odot$, despite their different internal structures and possibly different dynamos. In contrast,  plotting helicity density against poloidal energy density appears to separate stars into two families. At a given helicity density, the stars with mass below $0.5$M$_\odot$ appear to have excess poloidal energy density. Is it possible that they have an excess of poloidal field that does not link with the toroidal field and so does not contribute to the helicity? Alternatively, as the differential rotation rate is very low for these stars they might be covered with randomly oriented small scale fields that are not organised at large-scales. Such small-scale fields would not contribute to the helicity, but might explain the excess of poloidal field. In order to answer these questions we would need to map the variation in the fields of the lowest mass stars through their cycles, but this has not yet been done. Indeed, it is not yet clear on what timescale these stars may show cycles, if at all.

Given the apparent link between energetic and eruptive phenomena on the Sun with its magnetic helicity it is interesting to see the decline in helicity density with age (Fig. \ref{fig:hvsage}). This suggests active phenomena in stars may decline in time. Unfortunately, as we do not have ages for the lowest-mass stars in our sample, we cannot investigate the possible role of bifurcation.


\section{Summary and conclusions}
\label{sec:conclusion}

In this paper we have derived a general expression for calculating the magnetic helicity density of any star given its poloidal and toroidal magnetic field components and radius. Subsequently, we presented solar helicity densities along with the first helicity densities at the surfaces of 51 stars other than the Sun. Our main results are as follows, and all refer to the absolute average helicity density across the visible hemisphere only: 
 
 \begin{itemize} 
    \item The helicity density rises and reaches a plateau with decreasing stellar mass. The saturation occurs at $\sim$ 0.5 M$_\odot$. This is the result of the corresponding variation of the toroidal field.
    \item The helicity density rises with decreasing Rossby number R$_{\rm{o}}$ and reaches a maximum at R$_{\rm{o}}\sim$ 0.1. At lower Rossby numbers there is some evidence of a subsequent decrease.   
    \item For our mass range of $0.1-1.34$ M$_\odot$, the helicity density relates to the toroidal magnetic energy density according to $|\langle{h\,}\rangle|$ $\propto$ $\langle{\rm{B}^2_{\rm{tor}}\,\rangle}^{0.86\,\pm\,0.04}$ with a scatter consistent with the Sun's variation through its cycle.
    \item The variation of the helicity density with the poloidal energy density separates the lower- and higher-mass stars into two families indicated by different slopes similar to the results of \citet{See2015}.
    \item There is an overall decay of helicity density with age. 
 \end{itemize}

We conclude that the helicity density of stars with masses lower or higher than 0.5 M$_\odot$ are different if plotted against stellar mass, Rossby number or poloidal magnetic energy. The fact that the helicity density of a star of any mass can be determined by the strength of its toroidal magnetic field shows us that the change in behaviour for low-mass stars is due to their relatively strong poloidal fields \citep{See2015}.    

When comparing our stars to the Sun we find that, in terms of helicity density, the Sun appears to be a normal example of a star of its mass. Consequently, we suspect the spread in stellar helicity values (for M$_\star$ > 0.5 M$_\odot$) may be due to other stars undergoing cycles similar to the Sun, with their helicity density varying in time. We note however that \cite{Lehmann2019} showed a spread in values due to different stellar inclination angles, which may contribute to the scatter in stellar helicity values. In the future, given magnetic maps of the same star over a longer time period, we will investigate how its stellar helicity density evolves in time, compared to the solar case.

\section*{Acknowledgements}

The authors thank the referee, Prof Mitch Berger, for helpful comments. KL acknowledges financial support from the Carnegie Trust. MJ acknowledges support from STFC consolidated grant number ST/R000824/1. LTL acknowledges support from the Scottish Universities Physics Alliance (SUPA) prize studentship and the University of St Andrews Higgs studentship. DHM would like to thank both the UK STFC and the ERC (Synergy grant: WHOLE Sun, grant Agreement No. 810218) for financial support. VS acknowledges funding from the European Research Council (ERC) under the European Unions Horizon 2020 research and innovation programme (grant agreement No. 682393 AWESoMeStars). JFD and AAV acknowledge funding from the European Research Council (ERC) under the H2020 research and innovation programme (grant agreement 740651 NewWorlds and 817540, ASTROFLOW).




\bibliographystyle{mnras}
\bibliography{Paper/refs} 







\bsp	
\label{lastpage}
\end{document}